\documentclass{article}
\usepackage{spconf}
\usepackage{amsmath, amssymb, amsfonts}
\usepackage{graphicx}
\usepackage{hyperref}
\usepackage{url}
\usepackage{textcomp}
\usepackage{xcolor}
\usepackage{siunitx}
\usepackage{comment}
\usepackage{tabularx}
\usepackage{longtable} 
\usepackage{colortbl}
\usepackage{multirow, multicol, array, makecell}
\usepackage{booktabs, stfloats}
\usepackage{subcaption}
\usepackage{cleveref}
\usepackage{algorithmic}
\usepackage{enumitem}
\usepackage{bm}
\usepackage{flushend}

\crefname{section}{Section}{Sections}
\crefname{figure}{Figure}{Figures}
\crefname{table}{Table}{Tables}

\usepackage[style=ieee,mincitenames=1]{biblatex}
\AtEveryBibitem{\clearfield{doi}}

\usepackage{glossaries}
\newacronym{tts}{TTS}{Text to Speech}
\newacronym{stft}{STFT}{Short-Time Fourier Transform}
\newacronym{vc}{VC}{Voice Conversion}
\newacronym{cnn}{CNN}{Convolutional Neural Network}
\newacronym{roc}{ROC}{Receiver Operating Characteristic}
\newacronym{eer}{EER}{Equal Error Rate}
\newacronym{auc}{AUC}{Area Under the Curve}
\newacronym{msa}{MSA}{Multi-Head Self-Attention}
\newacronym{mlp}{MLP}{Multi-Layer Perceptron}
\newacronym{bce}{BCE}{Binary Cross Entropy}
\newacronym{mse}{MSE}{Mean Squared Error}

\makeatletter
\newcommand{\warn}[1]{\@latex@warning{Warn: #1}\textcolor{red}{WARN: #1}}
\newcommand{\todo}[1]{\@latex@warning{TODO: #1}\textcolor{red}{TODO: #1}}
\makeatother

\title{Multi-task Transformer for Explainable Speech \\
Deepfake Detection via Formant Modeling}

\name{Viola~Negroni$^*$, Luca~Cuccovillo$^\dagger$, Paolo~Bestagini$^*$, Patrick~Aichroth$^\dagger$, Stefano~Tubaro$^*$
\thanks{This work was supported by the FOSTERER project, funded by the Italian Ministry of Education, University, and Research within the PRIN 2022 program, and by the news-polygraph project (grant no. 03RU2U151D), funded by the German Federal Ministry of Research, Technology and Space (BMFTR).
This work was partially supported by the European Union - Next Generation EU under the Italian National Recovery and Resilience Plan (NRRP), Mission 4, Component 2, Investment 1.3, CUP D43C22003080001, partnership on “Telecommunications of the Future” (PE00000001 - program “RESTART”).
This work was partially supported by the European Union - Next Generation EU under the Italian National Recovery and Resilience Plan (NRRP), Mission 4, Component 2, Investment 1.3, CUP D43C22003050001, partnership on ``SEcurity and RIghts in the CyberSpace’’ (PE00000014 - program ``FF4ALL-SERICS’’).}}
\address{$^*$Dipartimento di Elettronica, Informazione e Bioingegneria - Politecnico di Milano - Milan, Italy\\
$^\dagger$Fraunhofer Institute for Digital Media Technology IDMT - Ilmenau, Germany}

\glsdisablehyper
\begin{document}
\ninept
\maketitle
\begin{abstract}

In this work, we introduce a multi-task transformer for speech deepfake detection, capable of predicting formant trajectories and voicing patterns over time, ultimately classifying speech as real or fake, and highlighting whether its decisions rely more on voiced or unvoiced regions. Building on a prior speaker-formant transformer architecture, we streamline the model with an improved input segmentation strategy, redesign the decoding process, and integrate built-in explainability. Compared to the baseline, our model requires fewer parameters, trains faster, and provides better interpretability, without sacrificing prediction performance.
\end{abstract}

\section{Introduction}
\label{intro}

We live in an era of rapid technological change, where AI has quickly become part of everyday life and a driving force across many fields. 
But this progress also carries serious risks: as society becomes more dependent on AI for routine tasks, and as AI tools become widely available, the chances of malicious use increase, often with harmful consequences.

This problem is especially evident on social media platforms, where
the rise of short, easily consumed content makes it easier for people to mistake synthetic media for real information. 
Among AI-generated media, \textit{deepfakes} refer specifically to synthetic content crafted to mislead or cause harm, and they are increasingly used in fraud, reputational attacks, and to spread misinformation online~\cite{amerini2025deepfake}.

Over the years, several data-driven models have been developed specifically for speech deepfake detection, and these detectors have laid a strong foundation in the field~\cite{li2025survey}.
However, the interpretability of these models remains a significant challenge, as it is difficult to determine in advance which speech features are most critical for their predictions of whether a given input is real or fake.
Furthermore, neural networks are known to exploit shortcuts during learning~\cite{geirhos2020shortcut, muller2021speech, negroni2024analyzing}. Even if this behavior is not inherently problematic, it highlights the importance of understanding what cues a model relies on when making decisions.
Indeed, several studies examining detector behavior have shown that many models base their decisions on unvoiced timesteps~\cite{sivaraman2025investigating}, frequency bands where no speech occurs~\cite{salvi2023towards}, or even background noise~\cite{salvi2024listening}, rather than the actual speech.
In all these cases, interpretability emerges only a posteriori, rather than being embedded in the model design itself.

Newer architectures like SFATNet~\cite{cuccovillo2023wifs, cuccovillo2024icassp, cuccovillo2024audio} have been proposed to improve interpretability by design. 
SFATNet models are hypothesis-driven, meaning their architecture should, in principle, encourage the network to focus on prosodic and speech-related cues. 
The latest version, SFATNet-3, achieves state-of-the-art performance, matching or surpassing established detectors such as RawNet2~\cite{tak2021end} and AASIST~\cite{jung2022aasist}.
Despite these advances, SFATNet-3 is computationally heavy due to its multi-layer transformer encoder-decoder structure. 
Moreover, while its outputs are designed to be interpretable, the model still lacks an explicit mechanism for self-explanation.

\begin{figure}[t]
    \centering
    \includegraphics[width=\columnwidth]{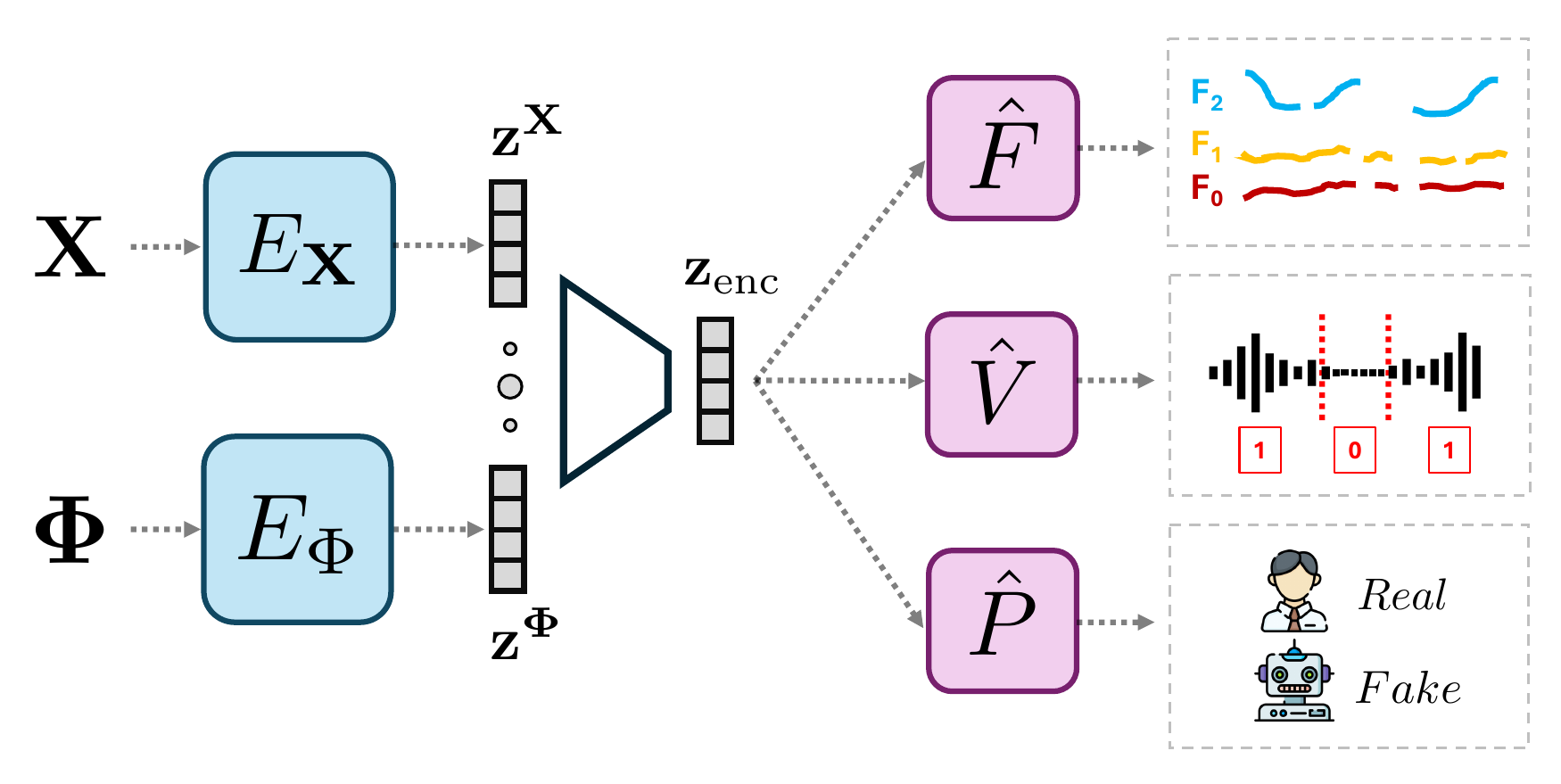}
    \caption{Illustrative example of the proposed multi-task transformer speech deepfake detector (SFATNet-4).}
    \label{fig:model}
\end{figure}

In this work we introduce SFATNet-4: a new, lightweight speech deepfake detector that is inherently and explicitly interpretable.
The proposed system is formulated as a multi-task model, where deepfake detection emerges as a byproduct of auxiliary objectives designed to enhance speech representation and prosodic awareness. 

\Cref{fig:model} provides an overview of the proposed architecture.
A transformer-based encoder, composed of two loosely coupled transformers, processes the magnitude and phase of the input speech signal separately, mapping them into a shared latent representation.
On the decoding side, each branch focuses on one task, working at the level of individual time frames.
One auxiliary branch encourages the network to learn patterns of prosodic variation by predicting the trajectories of fundamental frequency ($F_0$) and the first two formants ($F_1$ and $F_2$) across speech, while another completes this knowledge by learning to distinguish between voiced and unvoiced regions.
The third branch focuses on deepfake detection and is equipped with a pooling multi-head mechanism that assigns importance weights to each time frame. 
These weights not only highlight which segments of speech most influence the final prediction, but also reveal whether the model relied more heavily on voiced or unvoiced regions.

\section{Proposed system}
\label{system}

\subsection{Problem Formulation}
\label{subsec:problem}
The speech deepfake detection task is defined as follows. Given a discrete speech signal $\mathbf{x}$ with class $y \in \{0,1\}$ ($0$: real, $1$: fake), the goal is to build a detector $\mathcal{D}(\cdot)$ that outputs $\hat{y} = \mathcal{D}(\mathbf{x}) \in [0,1]$, the predicted probability of $\mathbf{x}$ being fake.

\subsection{Proposed Architecture}
\label{subsec:architecture}

\begin{figure}
    \centering
    \includegraphics[width=\columnwidth]{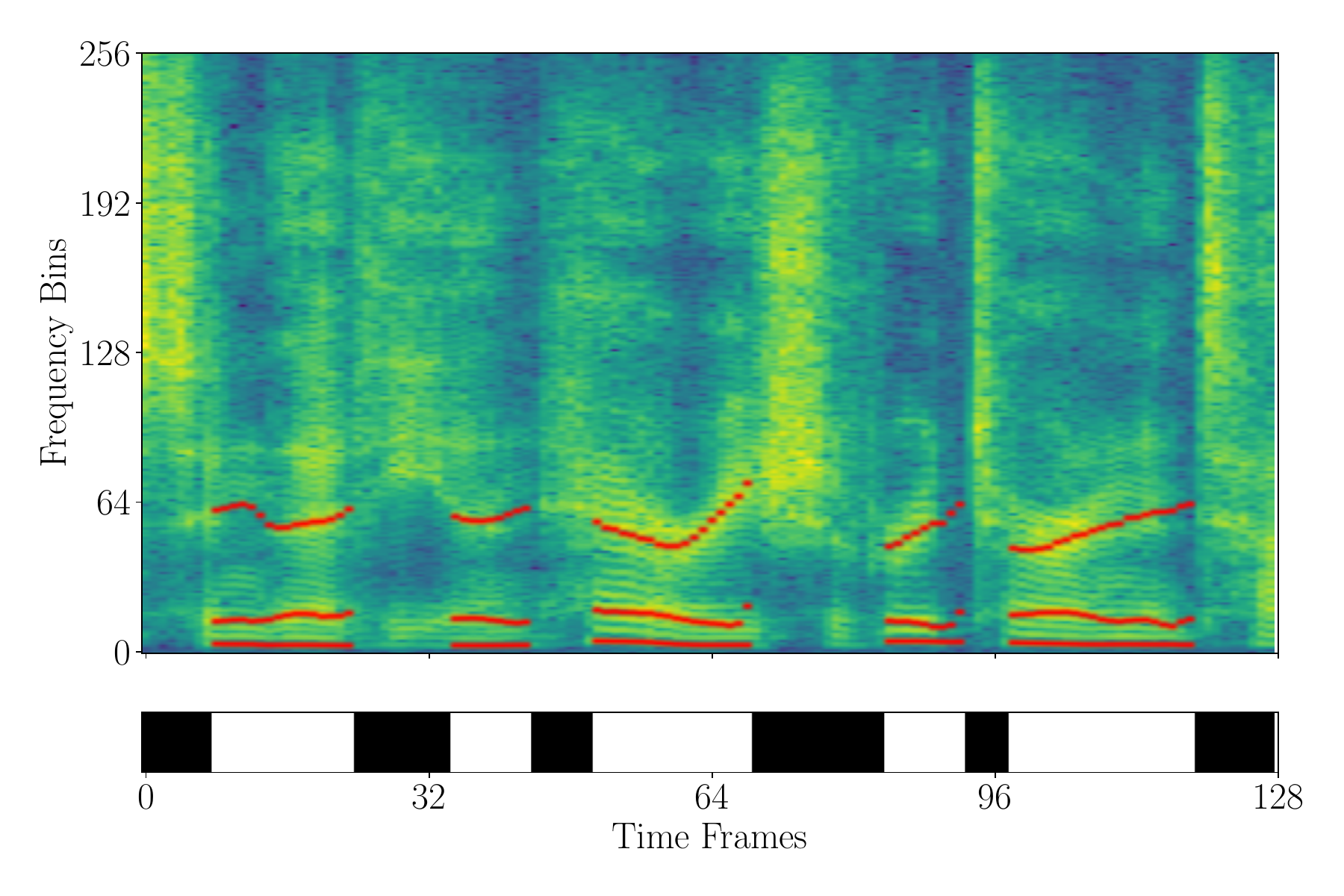}
    \caption{Example model output for an input spectrogram. Top: original spectrogram with predicted $F_0$, $F_1$, and $F_2$. 
    Bottom: predicted binary voiced/unvoiced segments (voiced: white, unvoiced: black). 
    }
    \label{fig:spec}
\end{figure}

The speech deepfake detector $\mathcal{D}(\cdot)$ that we introduce in this work is an enhanced version of the SFATNet-3 model~\cite{cuccovillo2024audio}, where we aim to improve its efficiency, computational weight, and interpretability.
Similar to its predecessor, the new detector is a multi-task audio transformer featuring two encoding modules, one for the magnitude and one for the phase of the input speech signal, and three decoding modules, each dedicated to a distinct task.

The Magnitude Encoder $E_\mathbf{X}(\cdot)$ and the Phase Encoder $E_\mathbf{\Phi}(\cdot)$ are transformer modules built from alternating layers of \gls{msa} and \gls{mlp} blocks, with layer normalization applied before each block and residual connections applied after, following the
standard transformer configuration~\cite{vaswani2017attention}. 
On the decoding side, the architecture includes: a Multi-formant Decoder $\hat{F}(\cdot)$ that predicts the evolution in time of the fundamental frequency $F_0$ and of the two formants $F_1$ and $F_2$, to capture prosodic patterns; a Voicing Decoder $\hat{V}(\cdot)$ that distinguishes between voiced and unvoiced regions; a Synthesis Predictor $\hat{P}(\cdot)$ dedicated to speech deepfake detection.

While $E_\mathbf{X}$ and $E_\mathbf\Phi$ remain identical to those in SFATNet-3, our model introduces three key changes. 
First, $\hat{P}(\cdot)$ was modified to enable improved explainability while retaining its original architecture.
Second, we completely redesigned $\hat{F}(\cdot)$ to enable more direct and fine-grained formant prediction. 
Third, we replaced the magnitude reconstruction task of the previous SFATNet versions with the prediction of voiced/unvoiced segments of speech by means of $\hat V(\cdot)$.
This new task becomes feasible in our architecture thanks to a revised segmentation strategy applied to the input \gls{stft}, a segmentation approach that enables a four-times faster training process. Changes are further detailed in the following.

\medskip

\noindent \textbf{Speech Preprocessing.} 
Let us denote the short-time Fourier transform of an input sample $\mathbf{x}$ by
\begin{equation}
X = \operatorname{STFT}(\mathbf{x}) \in \mathbb{R}^{L \times M},
\label{eq:stft}
\end{equation}
where $L$ is the number of frames and $M$ the number of frequency bins.  
From $X$, we compute the log-magnitude and phase representations as
$\mathbf{X} = \log(|X|)$ and $\mathbf{\Phi} = \sin(\angle X)$ respectively, maintaining a linear frequency axis.
These representations are then segmented to produce input tokens for $E_{\mathbf X}(\cdot)$ and $E_{\mathbf\Phi}(\cdot)$. 

In previous work, $\mathbf{X}$ and $\mathbf{\Phi}$ were segmented to obtain squared patches over both time and frequency as tokens. 
Our approach segments only along the time axis, producing slices of size $1 \times M$, i.e., a single time frame containing all frequency bins.
This time-only segmentation  not only reduces complexity and enables efficient processing, but also eventually allows for frame-level interpretability analysis.

\medskip
\noindent \textbf{Encoding Step.} 
The magnitude and phase tokens are linearly projected to $\mathbb{R}^{L \times D}$ and processed by $E_\mathbf{X}$ and $E_\mathbf{\Phi}$, 
producing
\begin{equation}
\mathbf{z}^\mathbf{X}, \mathbf{z}^\mathbf{\Phi} \in \mathbb{R}^{L \times D}.
\end{equation}
These sequences are concatenated ($\circ$) along the feature dimension $D$ and projected back to the original embedding size by a linear layer:
\begin{equation}
\mathbf{z}_\text{enc}  \in \mathbb{R}^{L \times D} = \left(\mathbf{z}^\mathbf{X}\circ\mathbf{z}^\mathbf{\Phi}\right)W_\text{enc}^\top + \mathbf{1}_L \cdot b_\text{enc}^\top,
\end{equation}
with $W_\text{enc}\in\mathbb{R}^{D \times 2D}$ the projection matrix, $b_\text{enc}\in\mathbb{R}^{D \times 1}$ the bias vector, and  $\mathbf{1}_L\in\mathbb{R}^{L \times 1}$ a $L$-dimensional column vector of ones. 

\medskip
\noindent \textbf{Multi-formant Decoder $\hat{F}(\cdot)$.} 
Formants are the resonant frequencies of the vocal tract that shape the sound of speech. They are influenced by the positioning and shape of the throat, mouth, and nasal passages as air flows through them.  
In SFATNet-3, the three primary formants ($F_0, F_1, F_2$) were predicted via a transformer module producing a matrix-like representation. 

In contrast, our model directly predicts the continuous trajectory of each formant, estimating its value for every input frame.
A linear projection maps the encoded tokens $\mathbf{z}_\text{enc}$ 
to a sequence $\mathbf{z}^{\hat F}$ of  three values per frame. These values are then converted to the final output $\hat{\mathbf{F}}$, representing the predicted formant trajectories across all $L$ frames by means of
\begin{equation}
   \hat{\mathbf{F}} \in \mathbb{R}^{L \times 3} = \left\lbrace\,\left(
   \sigma(\mathbf{z}^{\hat F}_i)\cdot F_i^+ + F_i^-
   \right)\,\right\rbrace_{ i\in\{1,2,3\}},
\end{equation}
with $\sigma(\cdot)$ the sigmoid function, and $F_i^+,F_i^-$ the upper and lower limit of the $i$-th formant. The ranges were selected according to physiologically plausible ranges, resulting in  $F_0 \in [60, 400]~\text{Hz}$, 
$F_1 \in [200, 850]~\text{Hz}$, and $F_2 \in [800, 2700]~\text{Hz}$.

This design removes the need for an additional transformer decoder for formants, and produces more interpretable frame-level predictions.
The top part of \Cref{fig:spec} illustrates the reconstructed formants of an example track on the original spectrogram, highlighting how the model effectively captures the speech structure.

\medskip
\noindent \textbf{Voicing Decoder $\hat{V}(\cdot)$.} 
This module replaces the magnitude reconstruction task in previous SFATNet versions by predicting voiced and unvoiced frames from the encoding embeddings $\mathbf{z}_\text{enc}$.  
Voiced frames are the ones in which vocal folds vibrate during speech production, whereas unvoiced frames correspond to sounds without vocal fold vibration and to non-speech segments.
A linear projection to $\mathbb{R}^{L \times 1}$ with sigmoid activation produces a probability for each frame, where frames with probability $\geq0.5$ are classified as voiced, as in the  bottom part of \Cref{fig:spec}, where the predictions of $\hat{V}(\cdot)$ over the example track are reported visually.
Importantly, the resulting binary mask $v_\text{mask} \in \{0,1\}^{L \times 1}$ is used to avoid formant prediction for unvoiced segments by refining the output of $\hat{F}(\cdot)$.

\medskip
\noindent \textbf{Synthesis Predictor $\hat{P}(\cdot)$.}
This module consists of a sequence-to-sequence transformer the output of which is pooled over time with a multi-head mechanism~\cite{martin2024exploring, xiao2025layer}, where tokens are scored via \textit{logsumexp} and softmax to produce frame-level weights.

Let us denote with $\mathbf{z}^{\hat P}\in\mathbb{R}^{L\times D}$ the raw output sequence of the transformer, given the input sequence $\textbf{z}_\text{enc}$. 
This raw output can be pooled by means of:
\begin{equation}
\mathbf{z}^{\hat P}_\text{pooled}\in\mathbb{R}^{1\times D}=\operatorname{softmax}\left(\log( \exp( \mathbf{z}^{\hat P} \cdot W_H) \cdot \mathbf{1}_H ) \right)^\top \cdot\mathbf{z}^{\hat P}, 
\end{equation}
with $W_H\in\mathbb{R}^{D\times H}$ the learned projection matrix for $H$ attention heads.
The pooled embedding can finally be converted into the synthesis prediction score $\hat{y}$ through layer normalization and a linear projection.

By replacing the standard class-token of earlier SFATNet versions with a pooling step, $\hat{P}(\cdot)$ is aligned with the other frame-level decoding modules, as it allows us to obtain one weight $w_l$ for each output element of the softmax layer, i.e., one attention score for each frame in the input sequence.
The pooling weights thus also reveal which frames drive the decision, enabling frame-level explainability that can be paired with the voicing predictor to assess whether the model relies more on voiced or unvoiced regions of speech.

\section{Experimental setup}
\label{setup}

\subsection{Datasets}
\label{subsec:datasets}
We consider \num{4} speech deepfake datasets to assess both in-domain performance and out-of-domain generalization.
All speech is resampled to $\SI{16}{\kilo\hertz}$ for consistency across datasets.

\begin{itemize}[leftmargin=*]
\item \textbf{ASVspoof 5~\cite{wang2025asvspoof}}.
Released for the 2024 edition of the homonymous challenge, this dataset was created through a crowdsourcing process, collecting data across a wide variety of acoustic environments. 
It features fake speech from \num{32} generators, a combination of both legacy and state-of-the-art \gls{tts} and \gls{vc} models.

\item \textbf{In-the-Wild~\cite{muller22_interspeech}}. 
This dataset is designed to evaluate speech deepfake detectors in realistic scenarios.
Audio clips from \num{54} celebrities and politicians were obtained by segmenting publicly accessible video and audio content.

\item \textbf{FakeOrReal~\cite{reimao2019dataset}}.
This dataset consists of both synthetic and real speech samples.
Fake samples were generated using \num{7} different open-source and commercial \gls{tts} systems.
Real speech samples were sourced from publicly available platforms.

\item \textbf{TIMIT-TTS~\cite{salvi2023timit}} A dataset of fake audio samples, generated from 12 different TTS methods. We paired it with its real counterpart, VidTIMIT~\cite{sanderson2002vidtimit}, and consider its \textit{clean} partition.
\end{itemize}
We train our model on ASVspoof 5 without any data augmentation, merging the \textit{train} and \textit{dev} partitions, splitting the resulting data 90/10, and oversampling real speech to maintain balanced classes.  
In-domain testing is performed on ASVspoof 5 \textit{eval} partition.
We also assess the model intrinsic generalization by evaluating it on three unseen datasets: In-the-Wild, FakeOrReal, and TIMIT-TTS.

\subsection{Training Specifications}
\label{subsec:training}

The ground truth annotations for $F_0$ were extracted using the pYin algorithm~\cite{mauch2014pyin}, a probabilistic pitch tracker that robustly estimates fundamental frequency trajectories. 
Since $F_0$ is only defined in voiced regions, frames where pYin yields a valid $F_0$ correspond to voiced segments, while frames without a pitch estimate can be classified as unvoiced. Hence, these $F_0$ annotations also provide supervision for the voicing decoder $\hat{V}(\cdot)$. 
This dual role ensures consistency across decoders: the Multi-formant Decoder $\hat{F}(\cdot)$ is supervised by continuous $F_0$ contours, while the Voicing Decoder $\hat{V}(\cdot)$ is trained with a binary voicing mask derived directly from the same ground truth signal, preventing contradictory predictions such such as unvoiced decisions in regions with nonzero pitch.

Higher-order formants ($F_1, F_2$) were obtained using Parselmouth’s Burg formant tracker~\cite{parselmouth,praat}.  
The frame and hop length we used are \num{0.032}$\,$s and \num{0.016}$\,$s, respectively. 
These values were also used for the STFT in the model’s preprocessing step, yielding $M = 256$ frequency bins and $L = 128$ time frames.  
In other words, the system processes fixed-length inputs of \SI{2.064}{\second}, corresponding to \num{33,024} samples. 
Shorter inputs were padded by repeating the signal, while longer utterances were truncated to match the required length.  
We removed leading and trailing silence and normalized the audio so that peak amplitudes reach \num{1.0}, to eliminate common shortcut artifacts that neural networks might otherwise exploit~\cite{wang2025asvspoof}.

As in SFATNet-3, the magnitude and phase encoders $E_X(\cdot)$ and $E_\Phi(\cdot)$ consist of \num{8} transformer layers, with \gls{msa} modules of \num{8} heads (dimension \num{64}) and \gls{mlp} blocks in \num{1024} dimensions. 
The synthesis predictor $\hat{P}(\cdot)$ also retains the same layout: \num{4} transformer layers, \gls{msa} with \num{6} heads (dimension \num{64}), and \gls{mlp} blocks in \num{1024} dimensions. 
The embedding dimension is $D = 512$ for all modules.
The new multi-head weighting mechanism in $\hat{P}(\cdot)$ has $H = 4$ heads.

The training used a batch size of \num{256} with the AdamW optimizer and an initial learning rate of \num{e-4}, which decays after \num{10} epochs of plateau. 
In contrast to SFATNet-3, no masking was applied to the encoders during training.
The model was trained for \num{100} epochs with early stopping (patience \num{20}), monitoring the validation loss.  
The compound loss combined a \gls{bce} loss for $\hat{P}$, a \gls{bce} loss for $\hat{V}$, and a \gls{mse} loss for $\hat{F}$, with weights \num{1}, \num{0.3}, and \num{0.3}, respectively.
To stabilize training, both target and ground-truth formants were first log-scaled and then standardized using means and standard deviations computed from the training set.  
During the \gls{mse} loss computation, only voiced frames are used to penalize the model.
On an NVIDIA A40 GPU, the model requires approximately \SI{15}{\minute} to train a single epoch, compared to over \SI{60}{\minute} for SFATNet-3. 
The number of parameters was reduced from \num{64.7}M to \num{41.8}M.

\section{Results}
\label{results}

\subsection{Performance and generalization}
\label{subsec:performance}

\begin{table*}[t]
\caption{Comparison of EER (\%) and AUC (\%) across datasets for the proposed model and SFAT-Net 3.}
\label{tab:results}
\centering
\begin{tabular}{ccc|cccccc|cc}
\hline
\toprule
\multirow{2}{*}{}    & \multicolumn{2}{c}{\text{ASVspoof 5}} & \multicolumn{2}{c}{\text{In-the-Wild}} & \multicolumn{2}{c}{\text{FakeOrReal}} & \multicolumn{2}{c}{\text{TIMIT-TTS}} & \multicolumn{2}{c}{\text{\textbf{Average}}} \\ 
                     \cmidrule(lr){2-3} \cmidrule(lr){4-5} \cmidrule(lr){6-7} \cmidrule(lr){8-9} \cmidrule(lr){10-11}
                     & EER $\downarrow$    & AUC $\uparrow$     & EER $\downarrow$     & AUC $\uparrow$     & EER $\downarrow$     & AUC $\uparrow$        & EER $\downarrow$     & AUC $\uparrow$   & EER $\downarrow$     & AUC $\uparrow$    \\ \midrule
Proposed             & \textbf{4.41}       & \textbf{98.89}     & \textbf{17.29}       & \textbf{89.17}     & \textbf{20.33}       & \textbf{85.03}        & 20.93                & \textbf{84.49}   & \textbf{15.74}       & \textbf{89.40}             \\
SFAT-Net 3           & 8.85                & 96.69              & 19.70                & 85.20              & 21.08                & 81.01                 & \textbf{18.59}       & 83.36            & 17.06                & 86.57             \\   \bottomrule  
\end{tabular}
\end{table*}

In \Cref{tab:results}, we compare the performance of our proposed model with its predecessor, SFATNet-3.
In-domain performance is evaluated on ASVspoof 5 (clean subset, no codecs), out-of-domain generalization is tested on In-the-Wild, FakeOrReal, and TIMIT-TTS.
Results are reported in terms of \gls{eer} and \gls{auc}.
Both models excel on the in-domain benchmark, with AUC values above 95\%.
Performance drops out-of-domain due to distribution shifts, as expected from omitting data augmentation, though both models remain reasonably robust.
Our proposed model outperforms SFATNet-3 on nearly all datasets, achieving lower \gls{eer} and higher \gls{auc}.
These results confirm the superiority of the proposed architecture both on seen data and unseen conditions.

\subsection{Robustness by design}
\label{subsec:robustness}

\begin{table}
\caption{Robustness of the proposed model across ASVspoof 5 codecs. Best in \textbf{bold}, worst in \textit{italic}.}
\label{tab:robustness}
\centering
\resizebox{\columnwidth}{!}{
\begin{tabular}{ccccccc}
\hline
\toprule
\multirow{2}{*}{}    & \text{Encodec} & \text{MP3} & \text{M4A} & \text{Opus} & \text{AMR} & \text{Speex}\\ \midrule
EER (\%)             & 29.2 & \textit{40.9} & \textbf{21.8} & 28.2 & 34.2 & 32.0 \\
AUC (\%)             & 77.5 & \textit{64.9} & \textbf{85.6} & 79.3 & 71.6 & 74.7 \\  
\bottomrule  
\end{tabular}
}
\end{table}

Robustness to common compression standards is essential for reliable deployment of spoofing detectors.  
In \Cref{tab:robustness}, we assess the intrinsic robustness (without training data augmentation) of our architecture by testing on codec-processed data from ASVspoof 5.
Results are broken down by codec type. 
Compared to in-domain results on clean data (\Cref{tab:results}), performances drop across all compression systems, the largest degradation being with MP3, and M4A showing higher resilience. 
When narrow-band (8kHz) data, available for Opus, AMR, and Speex, are considered performance drops further by 16.8\%, 12.0\%, and 14.7\% \gls{auc}, respectively.
Notably, the model maintains solid discriminative power under the newer neural codec Encodec.
Therefore, our model shows some inherent robustness, but data augmentation is required for reliable real-world performance.

\subsection{Explainability analysis}
\label{subsec:xai}

\begin{figure}[t]
\centering
\includegraphics[width=0.49\textwidth]{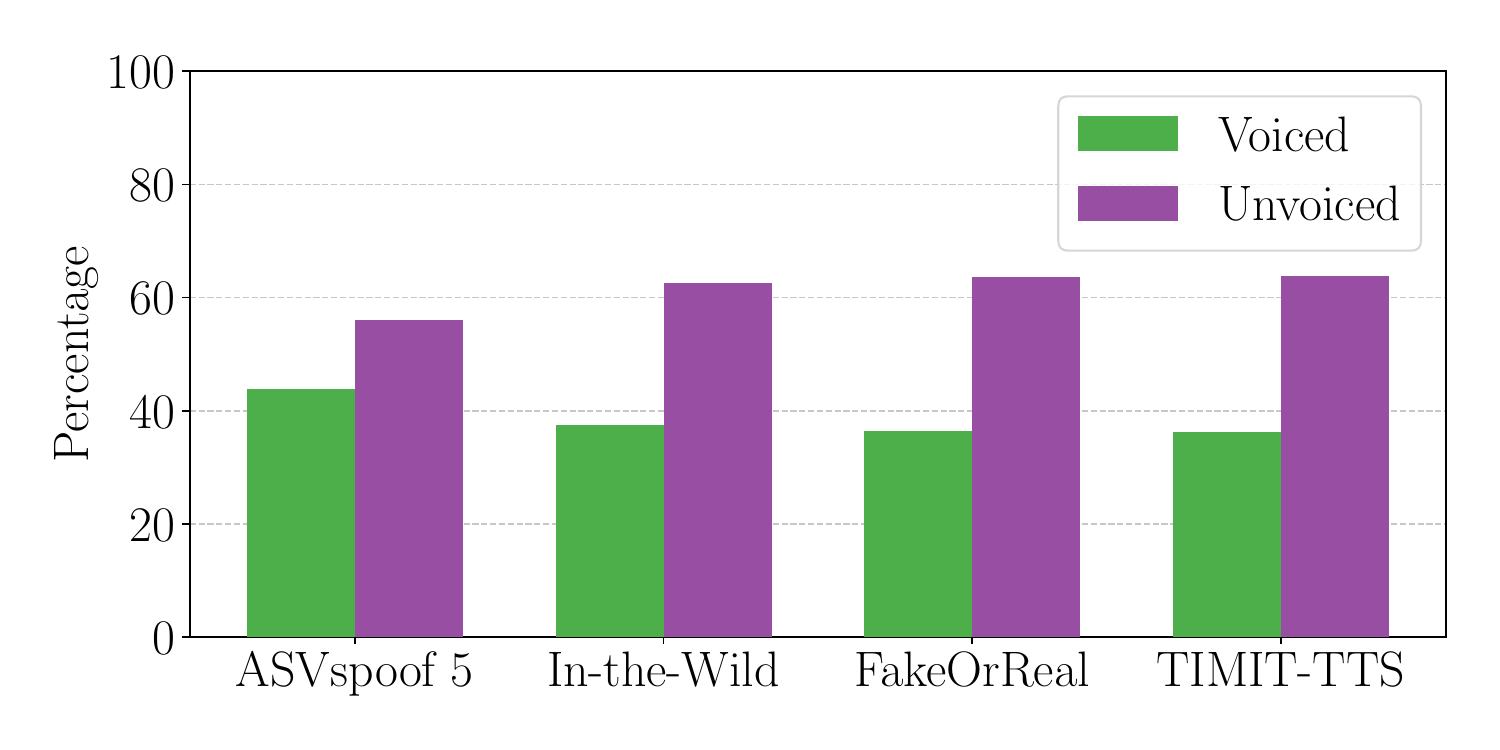} \quad
\includegraphics[width=0.49\textwidth]{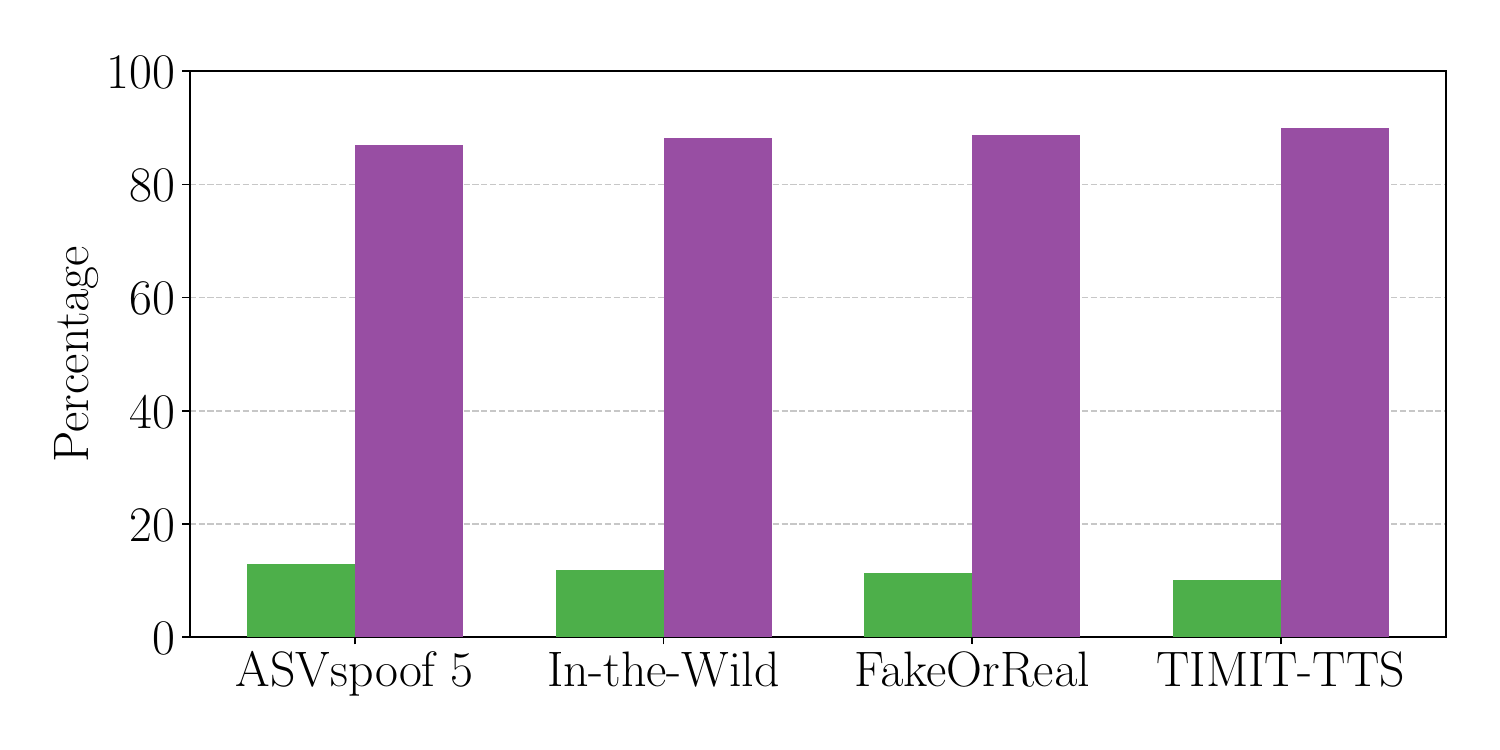} 
\caption{Relative importance of voiced and unvoiced frames for correctly classified speech. Each pair of bars shows the contributions of voiced and unvoiced frames to the model’s decision in a specific dataset, as weighted by the model’s attention scores. Top: real speech; bottom: fake.}
\label{fig:xai_analysis}
\vspace{-1.5em}
\end{figure}

The new input segmentation by time frames, combined with the pooling attention mechanism introduced in $\hat{P}(\cdot)$, enables \textit{temporal explainability} by highlighting the frames that contribute most to the model decisions.
When paired with the voicing predictor, also producing frame-level outputs, this makes it possible to determine whether the model relies more on voiced or unvoiced regions of speech for its predictions.
For this analysis, we focus on correctly predicted samples using the threshold corresponding to the model \gls{eer}, and we compute averages over all frame-level scores weighted by their attention values. 
Since the attention weights are very sparse in our case, this approach produces results that are essentially equivalent to a top-k study.
\Cref{fig:xai_analysis} presents the analysis of our model for the outcomes in \Cref{tab:results}.

For in-domain data from ASVspoof 5, the model shows a nearly equal reliance on voiced and unvoiced regions when processing real speech. For out-of-domain data, however, the model draws more cues from unvoiced parts of speech, which may correspond to speech segments without formants or to non-speech regions. 
Interestingly enough, for synthetic speech, the model consistently relies heavily on unvoiced regions.
This suggests that
the most salient synthesis artifacts are found in unvoiced segments, an observation that aligns with previous studies~\cite{sivaraman2025investigating}.

\section{Conclusions}
\label{conclusions}

In this work we presented SFATNet-4: a novel, lightweight multi-task transformer for speech deepfake detection. 
The model, designed for interpretability, simultaneously captures prosodic patterns, distinguishes voiced from unvoiced segments, and identifies manipulated speech.
The integration of attention-based mechanisms and auxiliary tasks not only enhances detection performance but also provides insight into the cues driving the model’s decisions, addressing a critical gap in current approaches.
Our results show that interpretability may be improved without sacrificing performance, paving the way for more transparent and accountable speech deepfake detection systems.
Future work will extend SFATNet-4 to multilingual speech, expand prosodic features, and optimize it for real-time use.

\clearpage
\section{References}
\printbibliography[heading=none]

\end{document}